\documentclass[aip, aps, %preprint, 
reprint,
amsmath,amssymb, floatfix
]{revtex4-1}
\usepackage{array}% http://ctan.org/pkg/array
\usepackage[per-mode=symbol,separate-uncertainty,range-phrase=--]{siunitx}
\usepackage{graphicx}% Include figure files
\usepackage{textcomp}
\usepackage{dcolumn}% Align table columns on decimal point
\usepackage{bm}% bold math
\usepackage{xcolor}

\usepackage{hyperref}
\hypersetup{
colorlinks = {true}, % color or boxlinks --- keep it true.
linkcolor = {black},
anchorcolor = {black},
citecolor = {black},
menucolor = {black},
filecolor = {black},
urlcolor = {blue},
runcolor = {black},
linkbordercolor = {black},
citebordercolor = {black},
urlbordercolor = {black},
menubordercolor = {black},
filebordercolor = {black},
runbordercolor = {black},
linktoc = {all},
pdfpagelayout = {SinglePage},
pdfstartview = {Fit},
unicode = {true},
}

\usepackage{lipsum}

\newcommand{\omr}{\ensuremath{\omega_\mathrm{r}}}
\newcommand{\gB}{\ensuremath{\gamma_\mathrm{T}}}
\newcommand{\TB}{\ensuremath{T_\mathrm{T}}}
\newcommand{\NB}{\ensuremath{N_\mathrm{T}}}

\newcommand{\gbar}{\ensuremath{\bar{\gamma}_\mathrm{T}}}
\newcommand{\kB}{\ensuremath{k_\mathrm{B}}}
\newcommand{\Fvect}{\ensuremath{\overrightarrow{F}}}
\newcommand{\gint}{\ensuremath{\gamma_\mathrm{x}}}
\newcommand{\Nint}{\ensuremath{N_\mathrm{x}}}
\newcommand{\gtr}{\ensuremath{\gamma_\mathrm{tr}}}
\newcommand{\Ntr}{\ensuremath{N_\mathrm{tr}}}

\newcommand{\Aalto}{QCD Labs, QTF Centre of Excellence, Department of Applied Physics, Aalto University, P.O. Box 13500, FI-00076, Aalto, Finland}
\newcommand{\TokyoMed}{College of Liberal Arts and Sciences, Tokyo Medical and Dental University, Ichikawa, 272-0827, Japan}
\newcommand{\UNSW}{Center for Quantum Computation and Communication Technology, School of Electrical Engineering and Telecommunications, University of New South Wales, Sydney, NSW 2052, Australia}
\newcommand{\Oulu}{Research Unit of Nano and Molecular Systems, University of Oulu, P.O. Box 3000, FI-90014,
Oulu, Finland}
\newcommand{\VTT}{VTT Technical Research Centre of Finland, QFT Center of Excellence, P.O. Box 1000, FI-02044, Aalto, Finland}

\begin{document}

\preprint{AIP/123-QED}

\title[]{Calibration of cryogenic amplification chains using normal-metal--insulator--superconductor junctions}% Force line breaks with \\

\author{E. Hyypp{\"a}}
 %\altaffiliation[Also at ]{Physics Department, XYZ University.}%Lines break automatically or can be forced with \\
 \affiliation{\Aalto}%

\author{M. Jenei}
\affiliation{\Aalto}%

\author{S. Masuda}%
 %\email{Second.Author@institution.edu.}
\affiliation{\TokyoMed}%
\affiliation{\Aalto}

\author{V. Sevriuk}
\affiliation{\Aalto}

\author{K. Y. Tan}

\affiliation{\Aalto}
\affiliation{\UNSW}

\author{M. Silveri}
\affiliation{\Aalto}
\affiliation{\Oulu}

\author{J. Goetz}
\affiliation{\Aalto}

\author{M. Partanen}
\affiliation{\Aalto}

%\author{\\J. Govenius}
%\affiliation{\Aalto}

\author{\\R. E. Lake}
\affiliation{\Aalto}

\author{L. Gr{\"o}nberg}
\affiliation{\VTT}

\author{M. M{\"o}tt{\"o}nen}
\affiliation{\Aalto}

%\date{\today}% It is always \today, today,
             %  but any date may be explicitly specified

\begin{abstract}
Various applications of quantum devices call for an accurate calibration of cryogenic amplification chains. To this end, we present a convenient calibration scheme and use it to accurately measure the total gain and noise temperature of an amplification chain by employing normal-metal--insulator--superconductor (NIS) junctions. 
Our method is based on the radiation emitted by inelastic electron tunneling across voltage-biased NIS junctions. 
We derive an analytical equation that relates  the generated power to the applied bias voltage which is the only control parameter of the device.
After the setup has been characterized using a standard voltage reflection measurement, the total gain and the noise temperature
are extracted by fitting the analytical equation to the microwave power measured at the output of the amplification chain. The 1$\sigma$ uncertainty of the total gain of \SI{51.84}{\decibel} appears to be of the order of \SI{0.1}{\decibel}. 
% * <mate.jenei@aalto.fi> 2018-06-22T08:34:50.205Z:
%
% ^.%Our method is based on the radiation emitted by electrons which tunnel inelastically across voltage-biased NIS junctions. 
%of the amplification chain 

%
%Valid PACS numbers may be entered using the \verb+\pacs{#1}+ command.
\end{abstract}

%\pacs{Valid PACS appear here}% PACS, the Physics and Astronomy
                            % Classification Scheme.
%\keywords{Suggested keywords}%Use showkeys class option if keyword
                              %display desired
\maketitle

Superconducting circuits provide a promising approach to implement a variety of quantum devices and to explore fundamental physical phenomena, such as the light-matter interaction \cite{wallraff2004strong} in the ultrastrong coupling regime \cite{niemczyk2010circuit}. In addition, superconducting circuits are potential candidates for building a large-scale quantum computer \cite{devoret2013superconducting, blais2004cavity}: superconducting qubits can be coupled in a scalable way \cite{majer2007quantumBus, sillanpaa2007twoBus,dicarlo2009twoqubit,dicarlo2010threeQubitEntanglement, brecht2016multilayer,barends2014threshold,corcoles2015demonstration,song201710}, and both the gate and the measurement fidelity of qubits exceed the threshold required for quantum error correction \cite{wang2011surface, barends2014threshold, kelly2015errorDetection,ofek2016extendingLifetime}.  

Since superconducting quantum circuits typically operate in the single-photon regime, signals are amplified substantially for readout \cite{mallet2009single,dewes2012characterization,devoret2013superconducting,lin2013single,riste2013deterministic,Heinsoo2018,Ikonen2018} using a chain of amplifiers, which is distributed over several temperature stages\cite{mallet2009single,dewes2012characterization}.
In the first stage, a near-quantum-limited amplifier \cite{macklin2015near}, such as a Josephson parametric amplifier \cite{yurke1988observation,yamamoto2008flux,castellanos2007widely,Pogorzalek_2016}, is often used to lower the noise temperature of the amplification chain \cite{kokkoniemi2018nanobolometer}. As a result of cascading several amplifiers, the uncertainty in the total gain of the amplification chain becomes significant and may complicate, for example, the estimation of the photon number in the superconducting circuit.  Therefore, accurate, fast, and simple methods for measuring the total gain of an amplification chain are desirable in the investigation of quantum electric devices. %CHECKED

%This is unfortunate since an accurate estimation of the total gain can be used to calculate the photon numbers present in the superconducting circuit.

The gain and the noise temperature of cryogenic amplifiers can be measured, for example, using superconducting qubits \cite{macklin2015near, Goetz_2016a}, Planck spectroscopy of a sub-kelvin thermal noise source \cite{mariantoni2010planck}, and the $Y$-factor method \cite{Yfactor, fernandez1998noise} which utilizes the Johnson--Nyquist noise emitted at different temperatures. In addition to these methods, shot noise \cite{blanter2000shot,spietz2003primary} sources, such as normal-metal--insulator--normal-metal junctions, can be used to determine the gain and noise temperature of cryogenic amplifiers \cite{chang2016noise,aassime2001radio}. However, this method typically requires a calibration measurement of the setup due to impedance mismatch\cite{chang2016noise}.  %CHECKED
\newpage
%The gain and noise temperature of cryogenic amplifiers can be measured, for example, using the standard method of hot and cold loads or the cryogenic attenuator method \cite{fernandez1998noise}. Both of these methods rely on measuring the output power of an amplifier for at least two different input noise powers, which enables the use of the so-called $Y$-factor technique \cite{Yfactor}. These standard methods yield rather accurate results but they are associated with a number of experimental challenges \cite{fernandez1998noise}, which include, for example, the knowledge of the accurate temperature profile of the measurement setup and relatively long thermalization times at low temperatures. In addition to these standard methods, shot noise \cite{blanter2000shot,spietz2003primary} sources, such as normal-metal--insulator--normal-metal (NIN) junctions, can be used to determine the gain and noise temperature of cryogenic amplifiers \cite{chang2016noise,aassime2001radio}. However, this method typically requires an additional calibration measurement of the setup due to impedance mismatch effects \cite{chang2016noise}.  
%\newpage

\begin{figure}[ht]
\centering
\includegraphics[width = 7.5cm]{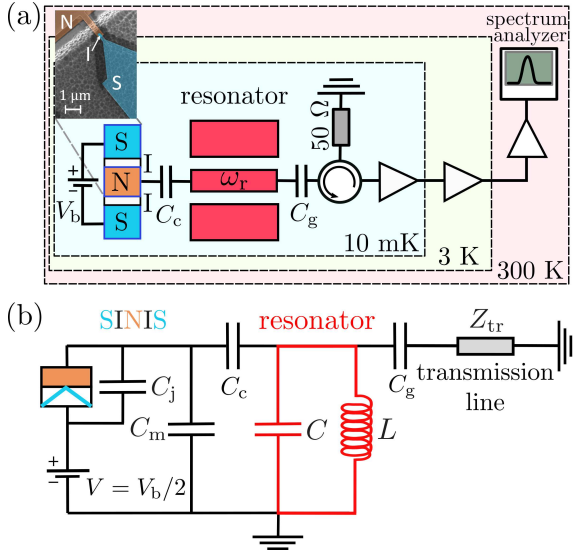}
\caption{ \label{fig: setup} (a) Schematic measurement setup. A bias voltage $V_\mathrm{b}$ is applied across an SINIS junction, which is capacitively coupled ($C_\mathrm{c}$) to a resonator. The bias voltage $V_\mathrm{b}$ tunes the photon emission rate of the junction and hence the mean photon number in the weakly coupled resonator\cite{tan2017quantum}. The other end of the resonator is capacitively coupled ($C_\mathrm{g}$) to a transmission line, which guides the output power to a spectrum analyzer. The inset shows a false-color scanning electron micrograph of a single NIS junction. (b) A simplified electric circuit diagram of the system depicted in (a). The capacitor $C_\mathrm{j}$ describes the capacitance  of a single NIS junction, whereas the capacitor $C_\mathrm{m}$ corresponds to the remaining capacitance. The fundamental mode of the resonator is modeled as an $LC$ oscillator. The characteristic impedance of the transmission line is $Z_\mathrm{tr} = 50 ~ \Omega$. }
\end{figure}

In this paper, we present an accurate alternative calibration scheme for the total gain and noise temperature of an amplification chain by utilizing photon-assisted electron tunneling in normal-metal--insulator--superconductor (NIS) junctions. To date, NIS junctions have been utilized in various applications, which include, for example, cryogenic microwave sources \cite{masuda2018observation},  thermometers \cite{kafanov2009single,gasparinetti2015fast}, and the recently developed quantum-circuit refrigerator that
cools quantum electric circuits by harnessing photon-assisted electron tunneling \cite{tan2017quantum,silveri2017theory,Silveri2018Lamb}. Here, we determine the gain and noise temperature of an amplification chain by measuring the power emitted by  electrons that tunnel inelastically across NIS junctions. The photon emission of the tunneling electrons can be activated by applying a bias voltage across the NIS junctions. For our analysis, we derive an analytic equation for the generated power in the high-bias regime. 
The analytic model matches our experimental results, which allows us to determine the gain of the amplification with an uncertainty of the order of \SI{0.1}\decibel. 
%Thus, the gain of an amplification chain can be conveniently determined by comparing the theoretical prediction with the measured power. We also demonstrate experimentally that our method can be used to calibrate the total gain of an amplification chain with an expected uncertainty of below 0.2 dB.  

We demonstrate the proposed calibration scheme on a sample illustrated in Fig.~\ref{fig: setup}. The device incorporates an SINIS junction which consists of two NIS junctions sharing a common normal-metal electrode. The normal-metal electrode of the tunnel junction is capacitively coupled to a half-wavelength superconducting coplanar-waveguide resonator. The resonator is further capacitively coupled to a transmission line from its other end, which conducts the signal to a three-stage amplification chain. %The transmission line power 
%Before detection, the emitted power is amplified using two amplifiers.
%The emitted power is amplified using a 38.5-dB cryogenic amplifier and a 20-dB  room temperature amplifier.
%which together with a 6-dB attenuator form the amplification chain under study. We also employ an isolator to prevent the noise generated by the amplifiers from reaching the CPW resonator. 
The sample is placed in a dry dilution refrigerator at 10-mK base temperature. Reference~\onlinecite{masuda2018observation} details the device fabrication. %\newpage

%Shorted version
%The samples are fabricated on a patterned 200-nm-thick Nb layer on top of a Si/SiO\textsubscript{2} wafer. By employing atomic layer deposition, the resonator is covered with a 50-nm-thick layer of Al\textsubscript{2}O\textsubscript{3} which serves as the dielectric material of the coupling capacitances. Subsequently, the NIS junctions are defined using electron beam lithography followed by two-angle electron beam evaporation and in-situ oxidation to introduce the metal electrodes (Al and Cu) and the insulating layer of the tunnel junctions. 

%Eric's version\
%The samples are fabricated on a high-purity 500-$\mu$m-thick Si wafer passivated with a 300-nm-thick layer of SiO\textsubscript{2}. The superconducting resonator is fabricated out of a 200-nm-thick Nb layer using photolithography and reactive ion etching. By employing atomic layer deposition, the resonator is covered with a 50-nm-thick layer of Al\textsubscript{2}O\textsubscript{3} which serves as the dielectric material of the coupling capacitances. Subsequently, the NIS junctions are defined using electron beam lithography followed by two-angle electron beam evaporation and in-situ oxidation to introduce the metal electrodes (Al and Cu) and the insulating layer of the tunnel junctions. 

\begin{figure}[ht]
\centering
\includegraphics[width = 8.4 cm]{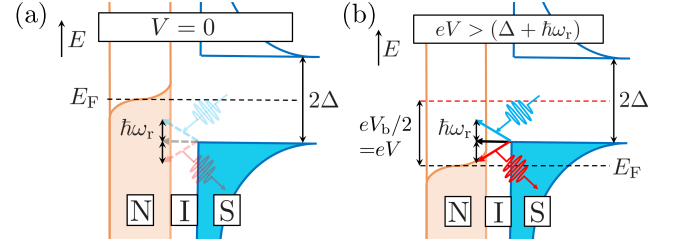}
\caption{ \label{fig: NIS tunneling} (a) Sketch of the occupied (shaded) and unoccupied (white) quasiparticle states in an NIS junction at vanishing bias voltage. In the normal metal (orange), the occupation is given by the Fermi--Dirac distribution since the density of states is essentially constant in the energy scale of interest. In the superconducting electrode (blue), there is an energy gap of $2 \Delta$ in the density of states, which restricts the possible electron tunneling events. Straight blue arrows indicate tunneling events involving photon (wavy arrow) absorption, whereas red arrows correspond to photon emission, and black arrows depict elastic tunneling events. The low transparency of the arrows highlights that all the above-mentioned tunneling events are suppressed.  (b) As~(a) but for a bias voltage $eV > \Delta + \hbar \omega_\mathrm{r}$ causing the Fermi level $E_\mathrm{F}$ of the normal metal (black dashed line) to shift with respect to that of the superconductor (red dashed line). For simplicity, we only show tunneling events originating from the lower edge of the superconductor gap. %This results in an exponential increase in the tunneling rates as compared with~(a).
}%CHECKED
\end{figure}

The bias voltage $V_\mathrm{b}$ across the SINIS junction activates the photon-assisted electron tunneling events which control the mean photon number $N_\mathrm{r}$ in the fundamental mode of the resonator. The photon number in turn determines the microwave power emitted to the transmission line. As described in Fig.~\ref{fig: NIS tunneling}, the electrons can tunnel through the NIS junction either elastically, i.e., without energy exchange with their electromagnetic environment, or inelastically by emitting or absorbing photons. In our setup, the resonator acts as the electromagnetic environment, and consequently the tunneling electrons absorb or emit photons at the resonance frequency  of the resonator $f_\mathrm{r} = \omega_\mathrm{r}/(2\pi)$ = 4.67 GHz. %CHECKED
%The CPW resonator, described by $f_\mathrm{r} = \omega_\mathrm{r}/(2\pi)$ fundamental mode frequency, constrains the photon energy absorbed or emitted by electron tunneling.
For vanishing bias voltage, both the elastic and inelastic tunneling events are suppressed due to the energy gap \cite{bardeen1957theory,dynes1984tunneling,giaever1960energy} of $2 \Delta$ in the superconductor density of states as shown in Fig.~\ref{fig: NIS tunneling}(a). If the bias voltage is slightly below the energy gap, i.e.,  $|eV_\mathrm{b}| \lesssim 2 \Delta$, electrons can tunnel through the junction by absorbing photons from the environment, which results in  cooling of the resonator mode. In this work, we are mostly interested in the high-bias-voltage regime $|eV_\mathrm{b}| \gg 2 \Delta$, where electron tunneling events involving photon emission are greatly enhanced, and hence the resonator mode heats up. The elevated temperature of the resonator mode leads to an increased radiative power into the transmission line, which enables us to calibrate the total gain of the amplification chain.% Due to the tunneling events, the electron temperature of the normal metal is also elevated at high bias voltages. In our analysis, we take into account the changing electron temperature only through its effect on electron tunneling since this is the dominating channel of heat flow.
%CHECKED
%Elastic tunneling is mode dominant always -> is this misleading
We show below that the power and the bias voltage relate to each other through a simple equation in the high-bias regime. 
We apply the theory developed in Ref.~\onlinecite{silveri2017theory} to describe the coupling between the resonator and the SINIS junction. In this model, we only take into account single-photon processes and we assume that the quasiparticle temperatures are equal in the normal-metal and superconducting electrodes. Furthermore, we assume sequential tunneling, i.e., that high-order processes are suppressed by the opaque tunnel barrier.
Using the simplified electric circuit in Fig.~\ref{fig: NIS tunneling}(b) and Fermi's golden rule, we can express the resonator damping rate $\gB$ and the effective temperature $\TB$ owing to the electron tunneling across the SINIS junction as \cite{silveri2017theory}
\begin{align}
\gB &= \gbar \frac{\pi}{\omr} \sum_{l,\tau = \pm 1 } l \Fvect (\tau eV + l \hbar \omr ), \label{eq: gamma_B} \\
\TB &= \frac{\hbar \omr}{\kB} \left \{ \ln \left [ \frac{\sum_{\tau = \pm 1} \Fvect (\tau eV + \hbar \omr)}{\sum_{\tau = \pm 1} \Fvect(\tau eV - \hbar \omr) } \right] \right \}^{-1}, \label{eq: TB}
\end{align}

where $\gbar$ is the asymptotic damping rate, $\Fvect(E)$ is the normalized rate of forward tunneling, $k_\mathrm{B}$ is the Boltzmann constant, $\hbar$ is the reduced Planck constant, and $V = V_\mathrm{b}/2$ is the voltage across a single NIS junction. We also have $\gbar = 2 C_\mathrm{c}^2 Z_\mathrm{r}\omr/[(C_\mathrm{c} + C_\mathrm{j} + C_\mathrm{m})^2R_\mathrm{T}]$, where $Z_\mathrm{r} = \sqrt{L/C}$, $R_\mathrm{T}$ is the tunneling resistance of a single NIS junction, and the remaining symbols are defined in Fig.~\ref{fig: setup}(b). 
The normalized rate of forward tunneling is defined as 
\begin{equation}
\Fvect(E) = \frac{1}{2\pi \hbar}\int \mathrm{d} \varepsilon ~ n_\mathrm{S}(\varepsilon)  \frac{f(\varepsilon - E) - f(\varepsilon)}{1 - \mathrm{e}^{-E/(\kB T_\mathrm{N})}},
\end{equation}
where $E$ is the energy gained by the tunneling electron, $T_\mathrm{N}$ is the temperature of the normal-metal electrode, $f(\varepsilon) = \{\exp[\varepsilon/(\kB T_\mathrm{N})] + 1]\}^{-1}$ is the Fermi function, and $n_\mathrm{S}(\varepsilon)$ is the Dynes density of states \cite{dynes1984tunneling}, which can be written as 
$n_\mathrm{S}(\varepsilon) = | \mathrm{Re} [ (\varepsilon + i \gamma_\mathrm{D} \Delta)/\sqrt{(\varepsilon + i \gamma_\mathrm{D} \Delta)^2 - \Delta^2} ] |$. The Dynes parameter $\gamma_\mathrm{D}$ describes the broadening of the superconductor energy gap, and is of the order of $\sim \! 10^{-4}$ in a typical experimental scenario \cite{pekola2010environment,tan2017quantum,masuda2018observation}.%CHECKED

In the high-bias regime, $eV \gg \Delta$, we employ Eqs.~\eqref{eq: gamma_B} and \eqref{eq: TB} to derive the following approximations for the damping rate $\gB$ and the effective photon number $\NB = \{\exp[\hbar \omr /(\kB \TB)] - 1\}^{-1}$ of the engineered environment
\begin{align}
\gB &\approx \gbar \left [ 1 + \frac{1}{2} \frac{\Delta^2}{(eV)^2} \right ] \label{eq: gamma_B high bias}, \\
\NB &\approx \frac{eV}{2 \hbar \omr} - \frac{1}{2} - \frac{\Delta^2}{2 \hbar \omr}\frac{1}{eV}, \label{eq: NB high bias}
\end{align}
where we have utilized the Sommerfeld expansion \cite{ashcroftsolid}. In addition, we have assumed that the Dynes parameter is small enough to be neglected at high bias voltages. %Note that the key equations \eqref{eq: gamma_B high bias} and \eqref{eq: NB high bias} were not presented in Ref.~\onlinecite{silveri2017theory}.
%CHECKED

The resonator exchanges energy with the SINIS junction and with the transmission line. Furthermore, the resonator may be subjected to additional sources of dissipation which we model as a single excess reservoir. Each of these three types of dissipation can be modeled as a virtual transmission line \cite{Goetz_2016b}, which allows us to write the net power flow between the resonator and the $i$th dissipative reservoir as
\begin{equation}
P_i = \hbar \omr \gamma_i (N_i - N_\mathrm{r}), \label{eq: power}
\end{equation}
where $\gamma_i$ is the damping rate of the resonator owing to the $i$th reservoir, $N_i$ is the corresponding effective photon number, and $N_\mathrm{r}$ is the resulting steady-state occupation of the resonator. Invoking the power balance and using Eqs.~\eqref{eq: gamma_B high bias} and \eqref{eq: NB high bias}, the net power flow into the transmission line can be approximated as
\begin{align}
P_\mathrm{tr} &\approx \frac{\gtr \gbar}{\gtr + \gbar + \gint} \bigg \{ \frac{eV}{2 } + \hbar \omr \bigg [ \frac{\gint (\Nint - \Ntr)}{\gbar} \nonumber \\
&-N_\mathrm{tr} - \frac{1}{2} \bigg] - \frac{1}{4} \frac{\Delta^2}{eV} \bigg( 1 + \frac{\gbar}{\gbar + \gtr + \gint} \bigg) \bigg \}, \label{eq: Ptr}
\end{align}
where $\gtr$ and $\Ntr$ are the damping rate and the effective photon number owing to the transmission line, respectively, whereas $\gint$ and $\Nint$ are the corresponding quantities for the excess losses.

\begin{figure*}
\centering
\includegraphics[width = 17 cm]{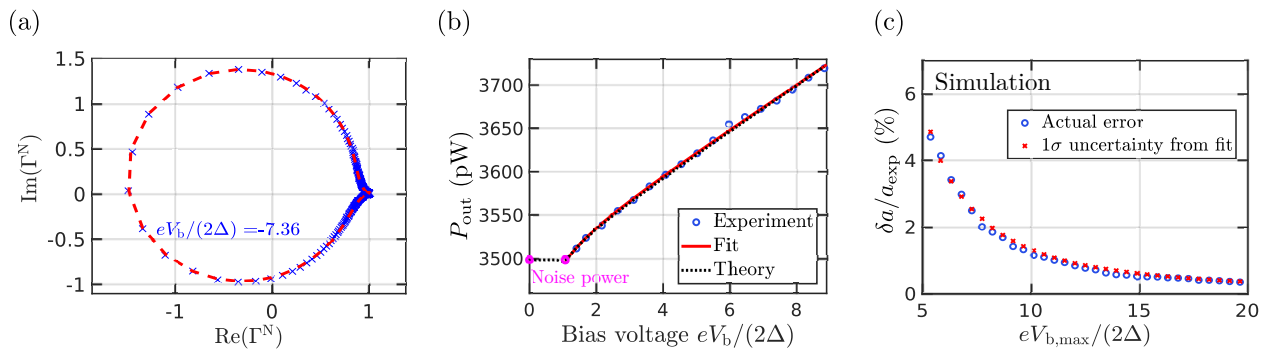}
\caption{\label{fig: results} (a) The normalized complex voltage reflection coefficient $\Gamma^{\rm{N}}$ (blue crosses) and the fit (red dashed line) that is obtained by a nonlinear regression where the ratio of two reflection coefficients of Eq. \eqref{eq: reflection measurement} is fitted.
(b)~The measured output power $P_\mathrm{out}$ as a function of the bias voltage (blue dots). The total gain $G$ is fitted using Eq. \eqref{eq: fitting function} (red line) in the voltage range $eV_\mathrm{b}/(2\Delta) \in [1.07, 8.83]$. The theoretical prediction for the power (dotted black line) is calculated by combining the fitted total gain, Eqs. \eqref{eq: gamma_B},  \eqref{eq: TB}, and \eqref{eq: power}. The magenta circles illustrate the noise power originating from the amplification chain. Each power data point is averaged over 100 repetitions. (c)~The average relative error of the fitting parameter $a$ (blue dots) and the average 1$\sigma$ uncertainty given by the fit (red crosses) as a function of the upper bound of the fitting range $eV_\mathrm{b}/(2\Delta) \in [1.07, eV_\mathrm{b, max}/(2 \Delta)]$ when Eq. \eqref{eq: fitting function} is fitted to 100 simulated power data sets. The simulated power data is obtained by adding normally distributed random noise to the exact theoretical prediction for the power calculated using Eqs. \eqref{eq: gamma_B},  \eqref{eq: TB}, and \eqref{eq: power}. The standard deviation of the random noise $\sigma = 2.5$ pW and the spacing between consecutive voltage points are chosen to coincide with the data shown in panel (b).}
\end{figure*}

In our experiments, we measure the output power of the amplification chain $P_\mathrm{out} = GP_\mathrm{tr} + P_\mathrm{noise}$, where $G$ is the total gain of the amplification chain including possible attenuation and losses, and $P_\mathrm{noise}$ is the noise power originating from the amplifiers. Consequently, we can determine the total gain $G$ by fitting a function of the form 
\begin{equation}
P_\mathrm{out}(V) = a V + b + c/V, \label{eq: fitting function}
\end{equation}
 to the measured power in the high bias regime $eV \gg \Delta$, where $\{a,b,c\}$ are the fitting parameters. Using Eq. \eqref{eq: Ptr}, the total gain $G$ can be expressed in terms of the leading-term coefficient as 
\begin{equation}
G = \frac{2 a}{e} \frac{\gbar +\gtr + \gint }{\gbar \gtr}. \label{eq: gain}
\end{equation}

Although we extract the total gain only from the coefficient $a$, the other terms in Eq. \eqref{eq: fitting function} improve the fit substantially. The effective noise temperature of the amplification chain $T_\mathrm{amp}$ is obtained by examining the output power at zero bias voltage $P_\mathrm{out}(0)$, where $P_\mathrm{tr}$ is practically zero with our device parameters and consequently
\begin{equation}
T_\mathrm{amp} \approx \frac{P_\mathrm{out}(0)}{G k_\mathrm{B} \Delta f},  \label{eq: noise temp}
\end{equation}
where $\Delta f$ is the bandwidth of the amplification chain.
%CHECKED
%%Jan's idea: right Sample A and Sample B in the figures.

In our experiments, we characterize the damping rates $\gbar$, $\gtr$, and $\gint$ with high accuracy leaving the parameter $a$ in Eq. \eqref{eq: gain} as the only free parameter in our model. To this end, we conduct standard microwave reflection measurements at different bias voltages $V_\mathrm{b}$. Based on the input-output theory \cite{gardiner1985input}, the voltage reflection coefficient of our system can be written as \cite{clerk2010introduction}
\begin{equation}
\Gamma = \frac{(2-r)\gtr - r(\gB + \gint) +2\mathrm{i}r(\omega_\mathrm{p} - \omr)}{\gB + \gtr + \gint - 2\mathrm{i}(\omega_\mathrm{p} - \omr)}, \label{eq: reflection measurement}
\end{equation}
where $\omega_{\rm{p}}/(2\pi)$ is the probe frequency and $r$ is a complex-valued Fano resonance correction factor, which arises from the direct cross-talk of the dissipative reservoirs\cite{fano2003temporal}.

Since the bias voltage $V_{\rm{b}}$ controls the coupling between the electromagnetic environment and the resonator, a Lamb shift arises for the resonance frequency\cite{Silveri2018Lamb} $\omr/(2\pi) = $ \SI{4.67}{\GHz}. The Lamb shift provides a convenient way of eliminating the unwanted background from the measurement data, namely, normalizing by the zero-bias measurement trace $\Gamma^{\rm{N}} (V_{\rm{b}}) = \Gamma (V_{\rm{b}})/\Gamma(0)$ as shown in Fig. \ref{fig: results}.(a). The ratio of two instances of Eq. \eqref{eq: reflection measurement} is fitted for every $V_{\rm{b}}$ above the critical coupling point, $eV_{\rm{b}}/(2\Delta)>1$, where the bias voltage-independent $\gtr$, $\gint$, and the bias voltage-dependent $\gB$ are fitting parameters. Next, the asymptotic damping rate is extracted using Eq. \eqref{eq: gamma_B high bias}. As a result, our characteristic damping rates are $\gtr/(2\pi) =$ \SI{1.78 \pm 0.02}{\MHz}, $\gbar/(2\pi) =$ \SI{17.39 \pm 0.04}{\MHz}, and $\gint/(2\pi) =$ \SI{0.46 \pm 0.01}{\MHz}, where the presented 1$\sigma$ uncertainties are obtained by the following method: First, an error circle is created that has a radius equal to the root mean square fit error and the center is located at $\Gamma^{\rm{N}}(V_{\rm{b}},\omr)$, which correspond to the resonance of the bias voltage-dependent feature. Finally, confidence intervals of each parameter are individually determined by finding the boundaries such that the fitted $\Gamma^{\rm{N}}$ is located within the error circle. The uncertainty of the excess damping rate is taken as the standard deviation of $\gint$ over the fitted voltage range.

After measuring the damping rates, we determine the total gain of the amplification chain by recording the microwave power at the output of the amplification chain for different bias voltages across the junction. To this end, we use a spectrum analyzer and numerically integrate the averaged spectral density around the resonance frequency $\omr /(2\pi)=4.67$ GHz in the range \mbox{4.6--\SI{4.75}{\GHz}}. 
%The generated transmission line power is obtained by calculating the average power spectral density at each bias voltage based on 21000 measurements and then numerically integrating the averaged spectral density around the resonance frequency $\omr /(2\pi) = 8.32$ GHz in the range 8.0 - 8.6 GHz. 
From the data shown in Fig.~\ref{fig: results}(b), we observe a monotonous increase of the microwave power if the bias voltage exceeds the energy gap. We finally extract the total gain of the amplification chain by fitting Eq. \eqref{eq: fitting function} to the power data. Using Eq. \eqref{eq: gain} and the experimentally determined device parameters, the total gain of the amplification chain is estimated to be $G = 51.84$~dB. %Tarkkaan ottaen virhearvio on 0.145 dB
%Using a transmission measurement  of a control sample, we arrive at roughly 39.5 dB, where the discrepancy between the measured gain arises from large uncertainties associated with the control measurement. 
Based on \mbox{Eq. \eqref{eq: noise temp}}, the average noise temperature of our amplification chain is \SI{11}{\kelvin} in the range 4.6--\SI{4.75}{\GHz}.
%CHECKED
 %The simulated results presented in Fig. \ref{fig: results}(c) do not, however, take into account the slight correlatedness of the observed noise in Fig. \ref{fig: results}(b), which may have a small effect on the uncertainty. 
 
%To estimate the achievable uncertainty of the total gain, we use the 1$\sigma$ uncertainty of the fitting parameter~$a$ obtained from Sample B ($\Delta a/a = 2$ \%) and assume that the relative errors of the damping rates coincide with those determined for Sample A using a reflection measurement. Based on the simulated results in Fig.~\ref{fig: results}(c), the 1$\sigma$ error bar of the fitting parameter $a$ is close to the actual error made in the fitting although Eq.~\eqref{eq: Ptr} is an approximation. Combining the uncertainties, we acquire $0.10$ dB for the 1$\sigma$ uncertainty of the total gain.  Since the damping rates of Sample B have not been extracted from a reflection measurement, this figure merely represents the performance of the calibration scheme in a typical setup. 
The relative uncertainty of the fitting parameter $a$ is extracted in the following way: we substitute the damping rates and the calibrated gain into Eq. \eqref{eq: gain} to determine the expected value $a_{\rm{exp}}$ for the fitting parameter $a$. 
We then create a set of simulated power spectra for bias voltages up to $V_{\rm{b}} = 10 \Delta / e$. Fitting Eq. \eqref{eq: fitting function} to the corresponding power values, we extract the parameter $a$ and record the absolute error $\delta a= a - a_{\rm{exp}} $ as a function of the maximum bias $eV_\mathrm{b,max}/(2\Delta)$. In addition, we obtain the $1\sigma$ uncertainty from the fit and observe that it agrees with the error $\delta a$. %Thus we may use the obtained caused by the simulated noisy power data presented in Fig.~\ref{fig: results}(c). 
In our experiments, we reach $eV_b / 2\Delta = 9$ and therefore estimate the relative uncertainty of the fitting parameter $\delta a/a_{\rm{exp}}= $ \SI{2}{\percent}.
%To estimate the achievable uncertainty of the total gain, use the 1$\sigma$ uncertainty of the fitting parameter~$a$ that is $\delta a/a = 2$ \%. Based on the simulated results in Fig.~\ref{fig: results}(c), the 1$\sigma$ error bar of the fitting parameter $a$ is close to the actual error made in the fitting although Eq.~\eqref{eq: fitting function} is an approximation.
Combining the uncertainties of the damping rates $\gbar$, $\gtr$, $\gint$, and the fitting parameter $a$, we acquire \SI{0.10}{\decibel} for the uncertainty of the total gain. 
%CHECKED

%In this paper, we have presented a simple measurement scheme for calibrating the total gain of a partly cryogenic amplification chain.
In this paper, we have presented a calibration scheme for the total gain and noise temperature of a general amplification chain that comprises cryogenic and non-cryogenic amplifiers. The knowledge of the gain and noise temperature is important when operating quantum devices and other low-power microwave components. 
%Knowledge of the gain is important, for example, when operating low-power electric devices. 
Currently, our setup allows the calibration of the total gain only at frequencies corresponding to a mode frequency of the resonator. The frequency range suitable for calibration can be expanded by placing a superconducting quantum interference device (SQUID) in the resonator \cite{partanen2018flux,sandberg2008tuning}. %which allows to tune the frequency of the resonator . 
In the future, we aim at benchmarking the accuracy of the proposed gain calibration scheme against a method utilizing a superconducting qubit in a resonator. By employing the ac Stark shift, we can determine the photon number in the resonator \cite{wallraff2004strong, schuster2005ac} and thus the total gain of the amplification chain.
%CHECKED

\begin{acknowledgments}
This project has received funding from the European Union's Horizon 2020 Research and Innovation Programme under the Marie Skłodowska-Curie Grant Agreement No. 795159 and under the European Research Council Consolidator Grant No.  681311 (QUESS), from the Academy of Finland Centre of Excellence in Quantum Technology Grant No.  312300 and No. 305237, from JST \mbox{ERATO} Grant No. JPMJER1601, from JSPS KAKENHI Grant No. 18K03486, and from the Vilho, Yrj{\"o} and Kalle V{\"a}is{\"a}l{\"a} \mbox{Foundation}.   
We acknowledge the provision of facilities and technical support by Aalto University at OtaNano - Micronova Nanofabrication Centre. 
%We thank \mbox{William} \mbox{Oliver}, \mbox{Greg} \mbox{Calusine}, \mbox{Kevin} \mbox{O'Brien}, and \mbox{Irfan} \mbox{Siddiqi} for providing us with the traveling-wave parametric amplifier used in the experiments.
\end{acknowledgments}
%CHECKED
%\nocite{*}
\bibliography{bibliography}% Produces the bibliography via BibTeX.

\end{document}